\documentclass{PoS}
\usepackage{psfig,epsfig}

\title{Improving perturbation theory with cactus diagrams}

\ShortTitle{Cactus Improvement}

\author{\speaker{Martha Constantinou} \\
        Univ of Cyprus, Physics Department, Nicosia, CY-1678, Cyprus\\
        E-mail: \email{phpgmc1@ucy.ac.cy}}

\author{Haralambos Panagopoulos\\
         Univ of Cyprus, Physics Department, Nicosia, CY-1678, Cyprus\\
        E-mail: \email{haris@ucy.ac.cy}}

\author{Apostolos Skouroupathis \\
        Univ of Cyprus, Physics Department, Nicosia, CY-1678, Cyprus\\
        E-mail: \email{php4as01@ucy.ac.cy}}

\abstract{We study a systematic improvement of perturbation theory for gauge fields
on the lattice~\cite{CPS}; the improvement entails resumming, to all orders in
the coupling constant, a dominant subclass of tadpole diagrams.

This method, originally proposed for the Wilson gluon
action~\cite{PV1}, is extended here to encompass all possible gluon
actions made of closed Wilson loops; any fermion action can be
employed as well.
The effect of resummation is to replace various parameters in the
action (coupling constant, Symanzik and clover coe{f}f{i}cient)
by ``dressed'' values; the latter are solutions to certain
coupled integral equations, which are easy to solve numerically.

Some positive features of this method are: a) It is gauge invariant, b) it
can be systematically applied to improve (to all orders) results
obtained at any given order in perturbation theory, c) it does indeed
absorb in the dressed parameters the bulk of tadpole contributions.

Two different applications are presented: The additive
renormalization of fermion masses, and the multiplicative renormalization
$Z_V$ ($Z_A$) of the vector (axial) current. In many cases where
non-perturbative estimates of renormalization 
functions are also available for comparison, the agreement with
improved perturbative results is consistently better as
compared to results from bare perturbation theory.}

\FullConference{XXIV International Symposium on Lattice Field Theory\\
		 July 23-28 2006\\
		 Tucson Arizona, US}

\begin{document}

\section{Introduction}

It is well known that quantities measured through numerical simulation
are characterized by significant renormalization effects, which must be
properly taken into account before making any comparisons to
corresponding physical observables.

Although the renormalization procedure can be formally carried out in
a systematic way to any given order in perturbation
theory, calculations are notoriously
di{f}f{i}cult, as compared to continuum regularization schemes.
Furthermore, the convergence rate of the resulting asymptotic series
is often unsatisfactory.

Some years ago, a method was proposed to sum up a whole subclass of
tadpole diagrams, dubbed ``cactus'' diagrams, to all orders in
perturbation theory~\cite{PV1,PV2}. This procedure is gauge invariant,
it can be systematically applied to improve (to all orders) results
obtained at any given order in perturbation theory, and it does indeed
absorb the bulk of tadpole contributions into an intricate
redefinition of the coupling constant. The agreement of available
non-perturbative estimates of renormalization coe{f}f{i}cients with
cactus improved perturbative results is consistently better as
compared to results from bare perturbation theory.

In the present work we extend the improved perturbation
theory method of Refs.~\cite{PV1,PV2}, to encompass the large class of
actions (including Symanzik improved gluon actions combined with any
fermionic action) which are used nowadays in simulations of QCD.
In Section II we present our calculation, deriving expressions for a
dressed gluon propagator. The methodology can be also applied to dress
the gluon and fermion vertices (appears in Ref.~\cite{CPS}). We show how these dressed
constituents are employed to improve 1-loop and 2-loop Feynman
diagrams coming from bare perturbation theory. In Section III we apply
our improved renormalizaton procedure to a number of test cases
involving Symanzik gluons and Wilson/clover/overlap fermions. 


\section{The Method}
\label{sec2}

In this Section, we start
illustrating our method by showing how the 
gluon propagator is dressed by the inclusion of cactus
diagrams. We will then explain how this procedure is applied to
Feynman diagrams at a given order in bare perturbation theory,
concentrating on the 1- and 2-loop case.

\subsection{Dressing the propagator}
We consider the Symanzik improved gluon action involving 
Wilson loops with up to 6 links:

\begin{eqnarray}
S_G=\frac{2}{g_0^2} \,\,& \Bigg[ &c_0 \sum_{\rm plaquette} {\rm Re\,
    Tr\,}(1-U_{\rm plaquette})\,  
  +  \, c_1 \sum_{\rm rectangle} {\rm Re \, Tr\,}(1- U_{\rm
    rectangle}) \nonumber \\  
 & + & c_2 \,\sum_{\rm chair} {\rm Re\, Tr\,}(1- U_{\rm chair})
\, 
  +  \, c_3 \sum_{\rm parallelogram} {\rm Re \,Tr\,}(1-
U_{\rm parallelogram})\Bigg]\,
\label{gluonaction}        
\end{eqnarray}
The coe{f}f{i}cients $c_i$ satisfy  a normalization condition $c_0 + 8 c_1
+ 16 c_2 + 8 c_3 = 1$ which ensures the correct classical
continuum limit of the action.

The quantities $U_i$ ($i = 0 ({\rm plaquette}),1 ({\rm rectangle}),2
({\rm chair}),3 ({\rm parallelogram})$)
in Eq. (\ref{gluonaction}) are products of 
link variables $U_{x,\mu}$ around the perimeter of the closed
loop. Using the
Baker-Campbell-Hausdorff (BCH) formula, $U_i$ takes the form:
\vspace{-2mm}
\begin{equation} 
U_i = \exp\,\Bigl(i\, g_0\, F^{(1)}_i + i\, g_0^2\, F^{(2)}_i +
 i\, g_0^3\,F^{(3)}_i + {\cal O}(g_0^4)\Bigr)
\label{BCH}\end{equation}
\vspace{-.5mm}
where $F^{(1)}_i$ is simply the sum of the gauge fields on the links
of loop $i$, while
$F^{(j)}_i\ (j>1)$ are $j$-th degree polynomials in the gauge
fields, constructed from nested commutators.

\noindent
\begin{minipage}[h]{3.8in}
\ \ \ Let us  {\it define} the cactus diagrams which dress the gluon
propagator: These are tadpole diagrams which become disconnected
if any one of their vertices is removed; further, each
vertex is constructed solely from the $F^{(1)}_i$ parts of the action.
A diagrammatic equation for the dressed gluon propagator (thick line)
in terms of the bare propagator (thin line) and 1-particle irreducible (1PI)
vertices (solid circle) reads:
\end{minipage}\hskip0.1\textwidth
\begin{minipage}[h]{0.5in}
\medskip
{\centerline{\psfig{figure=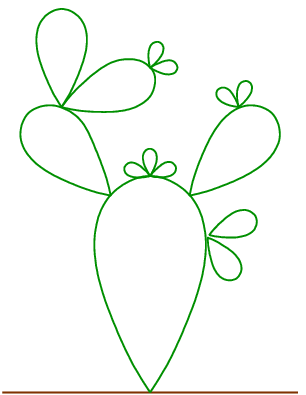,width=2.85truecm}}}
\medskip
{\centerline{Fig.1: A cactus}}
\end{minipage}
\medskip
\begin{equation}
\psfig{figure=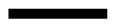,width=.7truecm}
= \psfig{figure=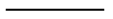,width=.7truecm} +
\psfig{figure=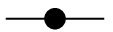,width=.7truecm} +
\psfig{figure=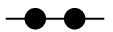,width=.7truecm} + \cdots
\label{propdress}
\end{equation}
 The 1PI vertex obeys the following recursive equation:
\begin{equation}
\psfig{figure=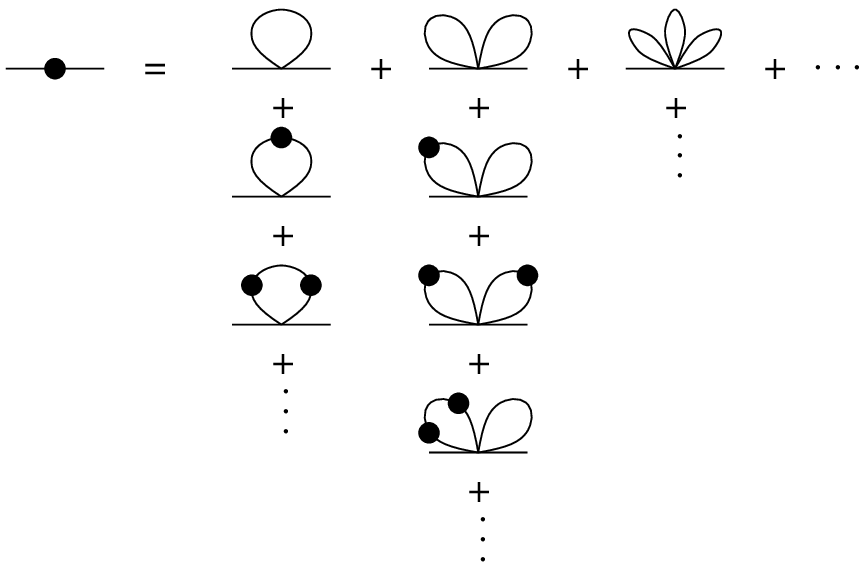,width=7truecm}
\label{recursive}
\end{equation}
The bare inverse gluon propagator $D^{-1}$ results 
from the total gluon action~(\cite{CPS}) and reads
\begin{equation}
\displaystyle D^{-1}_{\mu\nu}(k) = \sum_\rho \left( 
\hat{k}_\rho^2 \delta_{\mu\nu} - \hat{k}_\mu\hat{k}_\rho \delta_{\rho\nu} 
\right)  \, d_{\mu\rho} + {{\hat{k}_\mu\hat{k}_\nu}\over 1-\xi} \equiv
\sum_{i=0,1,2,3} \left( c_i\,G^{(i)}_{\mu\nu}(k) \right)
 + {{\hat{k}_\mu\hat{k}_\nu}\over 1-\xi}
\end{equation}
where  $C_0 = c_0 + 8 c_1 + 16 c_2 + 8 c_3, \quad C_1 = c_2 +
c_3, \quad C_2 = c_1 - c_2 - c_3 $. 
For further definitions of quantities appearing above, the reader can
refer to \cite{CPS}. The matrices $G^{(i)}(k)$ are symmetric and transverse, and
originate from a ${\rm Tr}\bigl(F^{(1)}_i\,F^{(1)}_i\bigr)$ term of the gluon action.
Consequently, the diagrams on the r.h.s. of
Eq.~(\ref{recursive}), are a
linear combination of $G^{(i)}(k)$; this implies that the 1PI vertex
$G^{\rm 1PI}(k)$ (the l.h.s. of Eq.~(\ref{recursive})) can be written as:
\begin{equation}
G^{\rm 1PI}(k) = \alpha_0\,G^{(0)}(k) 
+ \alpha_1\,G^{(1)}(k)
+ \alpha_2\,G^{(2)}(k) + \alpha_3\,G^{(3)}(k)
\end{equation}
Each of the quantities $\alpha_i$ will in general depend on $N,\ g_0,
\ c_0,\ c_1,\ c_2,\ c_3$, but not on the momentum. 
Eq.~(\ref{propdress}) leads to the following expression for the
inverse dressed propagator $(D^{\rm dr})^{-1}(k)$ ~\cite{CPS}:
\begin{eqnarray}
(D^{\rm dr})^{-1} = \tilde c_0\,G^{(0)} + \tilde c_1\,G^{(1)} +
\tilde c_2\,G^{(2)} +\tilde c_3\,G^{(3)} + 
{1\over 1-\xi}\, \hat{k}_\mu\hat{k}_\nu\,,\qquad 
\tilde c_i\equiv c_i-\alpha_i\label{Ddressed}
\end{eqnarray}

We observe that dressing replaces the bare coe{f}f{i}cients $c_i$ with improved ones
$\tilde c_i$\,, and leaves the longitudinal part intact. This property
ensures gauge invariance of the results.

In terms of the dressed propagator, Eq.~(\ref{recursive}) can be {\it
  drawn} as:
\begin{equation}
\psfig{figure=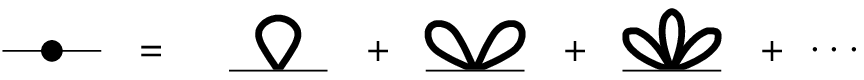,width=8truecm}
\label{recursive2}
\end{equation}
A typical diagram on the r.h.s. of
Eq.~(\ref{recursive2}) is the sum of 4 terms, and has
 $(n-2)/2$ 1-loop integrals in the diagram (coming from the contraction of two powers of
$F^{(1)}_i$ via a dressed propagator), and will contribute one power of
$\beta_i(\tilde c_0, \tilde c_1, \tilde c_2, \tilde c_3)$, where: 
\begin{eqnarray}
\beta_0 &=&\int_{-\pi}^\pi \frac{d^4q}{(2\pi)^4} \,
\Bigl(2\,\hat q_\mu^2\,D^{\rm dr}_{\nu\nu}(q) - 2\, \hat q_\mu\,
\hat q_\nu\,D^{\rm dr}_{\mu\nu}(q)\Bigr)\nonumber\\
\beta_1 &=&\int_{-\pi}^\pi \frac{d^4q}{(2\pi)^4} \,
\Bigl((4\hat q_\nu^2 - \hat q_\nu^4)\,D^{\rm dr}_{\mu\mu}(q)+
      \hat q_\mu^2 (4{-} \hat q_\nu^2)\,D^{\rm dr}_{\nu\nu}(q) 
      -2\,\hat q_\mu\,\hat q_\nu (4{-} \hat q_\nu^2)\,
D^{\rm dr}_{\mu\nu}(q)\Bigr)\nonumber\\
\beta_2 &=&\int_{-\pi}^\pi \frac{d^4q}{(2\pi)^4} \,
\Bigl(\hat q_\mu^2 (8{-} \hat q_\nu^2)\,D^{\rm dr}_{\rho\rho}(q)/2
-\hat q_\mu\,\hat q_\rho (8{-} \hat q_\nu^2)\,
D^{\rm dr}_{\mu\rho}(q)/2\Bigr)
\nonumber\\
\beta_3 &=&\int_{-\pi}^\pi \frac{d^4q}{(2\pi)^4} \,
\Bigl(3\,\hat q_\mu^2 (4{-} \hat q_\nu^2)\,D^{\rm dr}_{\rho\rho}(q)/2
-3\,\hat q_\mu\,\hat q_\nu\, (4{-} \hat q_\rho^2)\,
D^{\rm dr}_{\mu\nu}(q)/2\Bigr)\label{beta}
\end{eqnarray}
($\mu, \nu, \rho$ assume distinct values; no summation implied).
We note that $\beta_i$ are gauge independent, since the
longitudinal part cancels in the loop contraction.

In order to set Eq.~(\ref{recursive}) in a mathematical form, we need to evaluate
$F(n;N)$ which is the sum over all complete pairwise
contractions of ${\rm Tr}\{T^{a_1} T^{a_2}\ldots T^{a_n}\}$.
Use of $F(n;N)$, along with the integrals (\ref{beta}), allows us to
resum (\ref{recursive}), leading to~\cite{CPS}:
\begin{eqnarray}
{c_i{-}\alpha_i\over c_i}\,(N^2{-}1) = e^{-\beta_i\,g_0^2\,(N{-}1)/(4N)}\,\left({N{-}1\over N}\,L^1_{N-1}
(g_0^2\,\beta_i/2) + 2\, L^2_{N-2}(g_0^2\,\beta_i/2)\right)
\label{cimp}
\end{eqnarray}
($L^\alpha_\beta$(x): Laguerre polynomials). Eqs.~(\ref{cimp})
are 4 separate equations where unknown quantities are the coe{f}f{i}cients $\alpha_i$\,;
they appear on the l.h.s., as well as inside the integrals $\beta_i$ of
the r.h.s, by virtue of Eqs. (\ref{beta}, \ref{Ddressed}). It is worth mentioning that 
all combinatorial weights are correctly incorporated in the procedure.

Eqs.~(\ref{cimp}) can be solved numerically and each choice of values for 
($c_i\,,\,g_0\,,\,N$) leads to a set of values  for $\tilde c_i\equiv
c_i-\alpha_i$ that are no longer normalized;
one may express the results of our
procedure in terms of a normalized set of improved coe{f}f{i}cients,
$\tilde c_i/\tilde C_0$ and an improved coupling constant $\tilde
g_0^2 = g_0^2/\tilde C_0$, where: 
$\tilde C_0 = \tilde c_0 + 8 \tilde c_1 + 16 \tilde c_2 + 8 \tilde
c_3$\,.
For reasons of simplicity we define rescaled quantities:
\begin{equation}
\gamma_i \equiv {c_i\over g_0^2}\,,\qquad
\tilde\gamma_i \equiv {\tilde c_i\over g_0^2}\,,\qquad
\tilde\beta_i(\tilde c_0, \tilde c_1, \tilde c_2, \tilde c_3) \equiv
g_0^2 \,\beta_i(\tilde c_0, \tilde c_1, \tilde c_2, \tilde c_3) =
\beta_i(\tilde\gamma_0, \tilde\gamma_1, \tilde\gamma_2,
\tilde\gamma_3)
\end{equation}
$\tilde\gamma_i$ must now satisfy the
coupled equations:
\begin{equation}
\tilde\gamma_i = {1\over N^2{-}1}\, \gamma_i \,
e^{-\tilde\beta_i\,(N{-}1)/(4N)}\,\left({N{-}1\over N}\,L^1_{N-1}
(\tilde\beta_i/2) + 2\, L^2_{N-2}(\tilde\beta_i/2)\right)
\label{rescaled}\end{equation}
For $SU(2)$ and $SU(3)$ the Laguerre polynomials
have a simple form and Eqs.~(\ref{rescaled}) can be written explicitly ~\cite{CPS}.

Since Eqs.~(\ref{rescaled}) have the form $x = f(x)$, they can be numerically solved using a
fixed point procedure~\cite{CPS}. A unique solution for $\tilde\gamma_i$ always
exists for all physically interesting values of $c_i$, and for all
values of $g_0$ well inside the strong coupling region. The convergence of the
procedure has been verified in a number of extreme cases.


\subsection{Numerical values of improved coe{f}f{i}cients}
We now present the values of the dressed coe{f}f{i}cients for several
gluon actions of interest. In Figs. 2-5, one can see the improved
coe{f}f{i}cients for Plaquette, Tree-level Symanzik, Iwasaki and Tadpole
improved L\"uscher-Weisz actions. Results for DBW2 action
are listed in Table I. 
\newline

\hspace{-1.5cm}
\begin{minipage}[h]{3in}
\begin{center}
\epsfig{figure=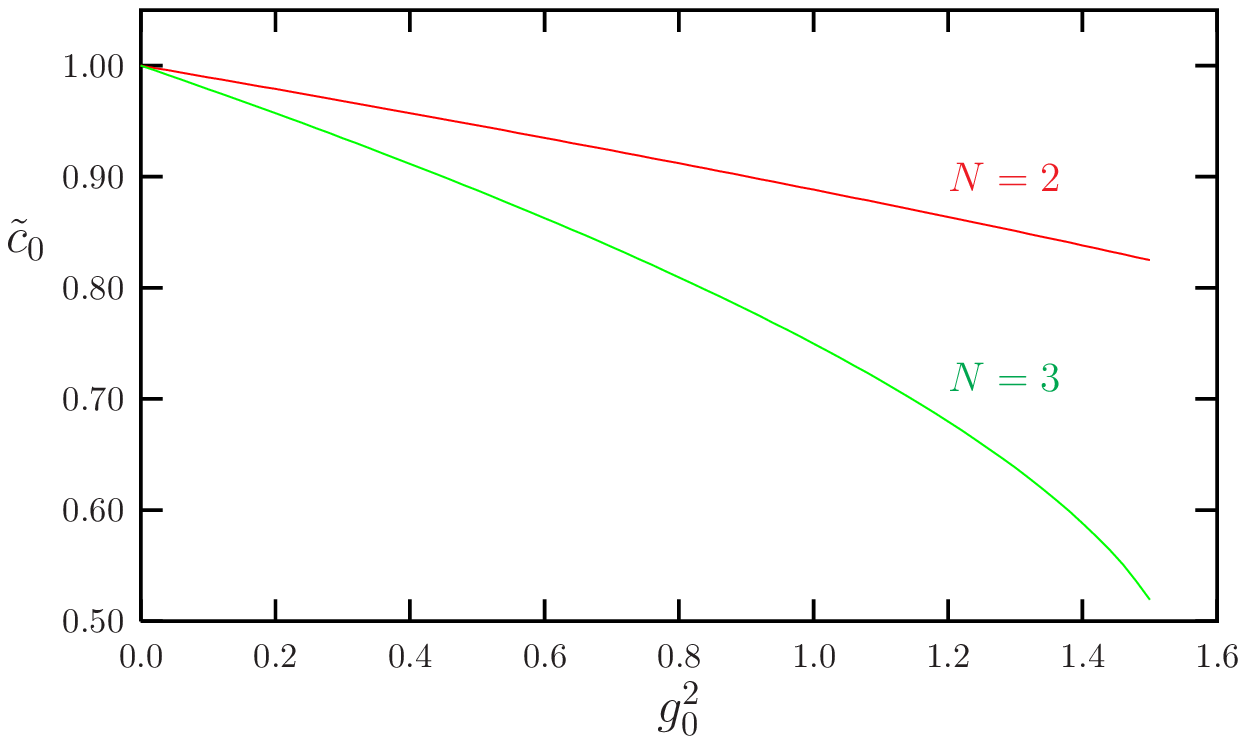,width=8truecm}
\vskip 1mm
\small{\label{fig2}Fig.2: Improved coe{f}f{i}cient\ \~{c}$_0$ for
  $N{=}2$ and $N{=}3$ (plaquette action)}
\end{center}
\end{minipage}
\hfill
\begin{minipage}[h]{3.5in}
\begin{center}
\epsfig{figure=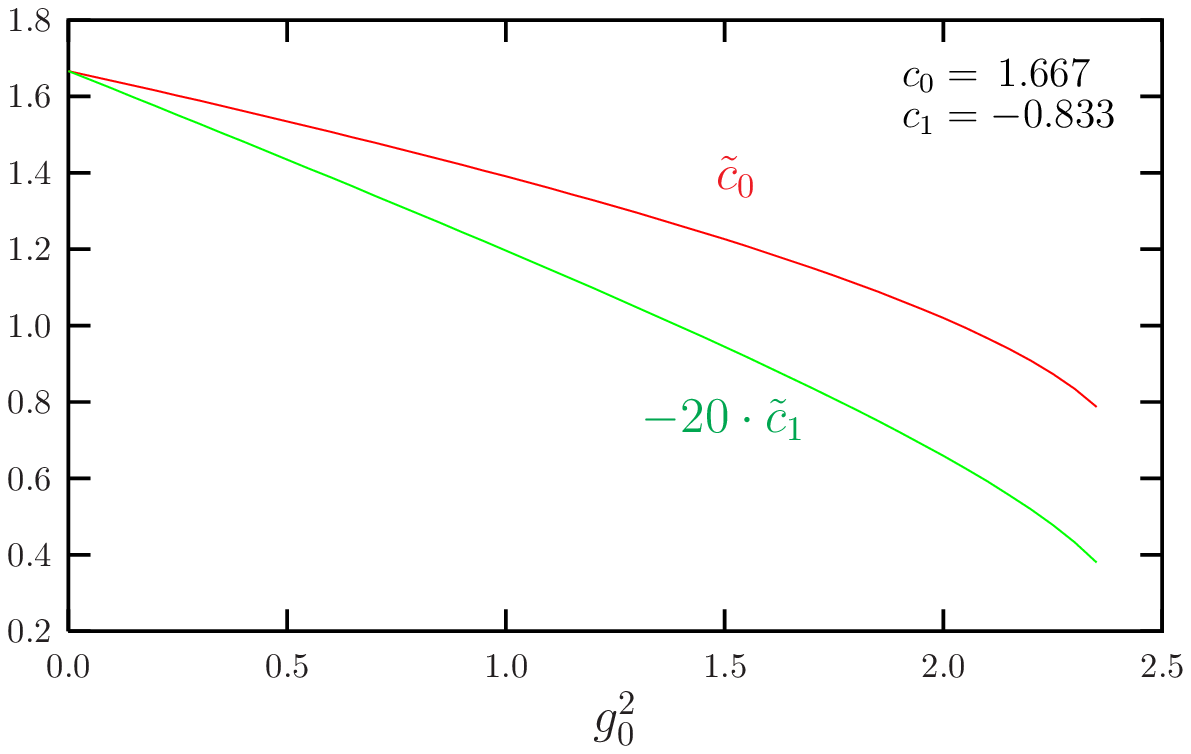,width=7.75truecm}
\vskip 1mm
\small{\label{fig3}Fig.3: Improved coe{f}f{i}cients\ \~{c}$_0$ and 
\ \~{c}$_1$ (tree-level \\ Symanzik improved action, $N=3$)}
\end{center}
\end{minipage}
\vskip 1mm
\vspace{5mm}
\hspace{-1.5cm}
\begin{minipage}[h]{3in}
\begin{center}
\epsfig{figure=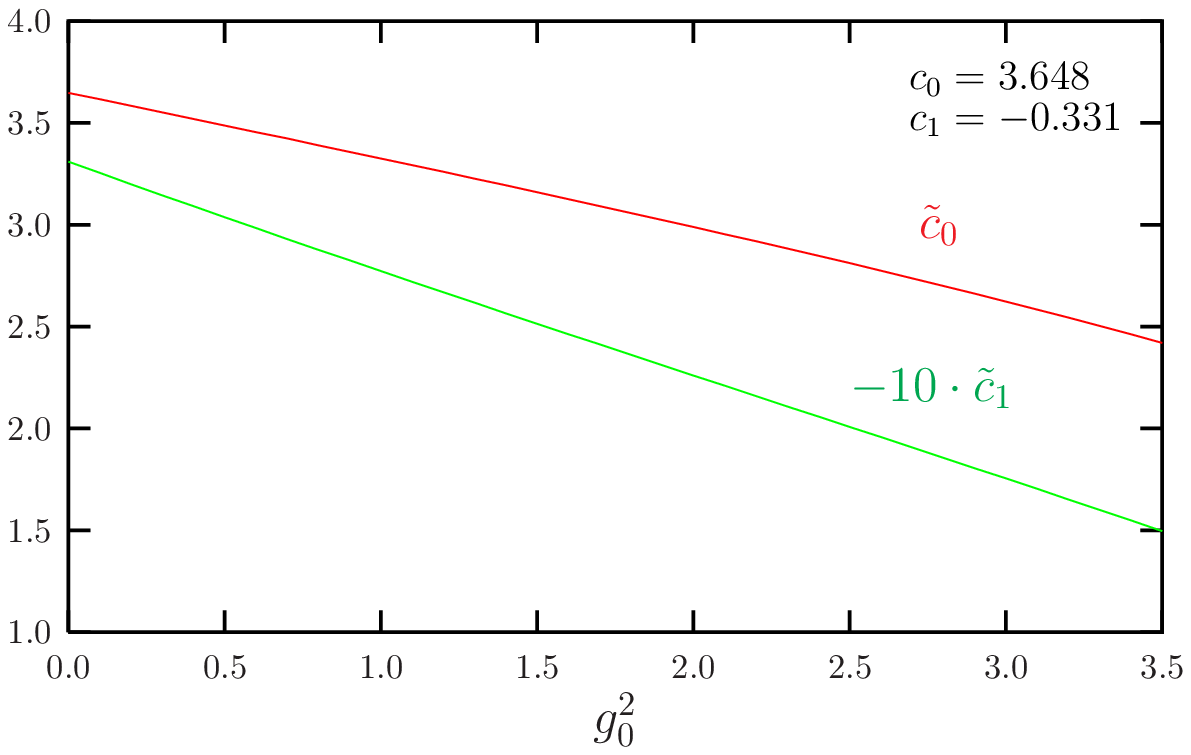,width=7.5truecm}
\vskip 1mm
\small{\label{fig4}Fig.4: Improved coe{f}f{i}cients\ \~{c}$_0$ and\ \~{c}$_1$ (Iwasaki action, $N=3$)}
\end{center}
\end{minipage}
\hfill
\begin{minipage}[h]{3.5in}
\begin{center}
\epsfig{figure=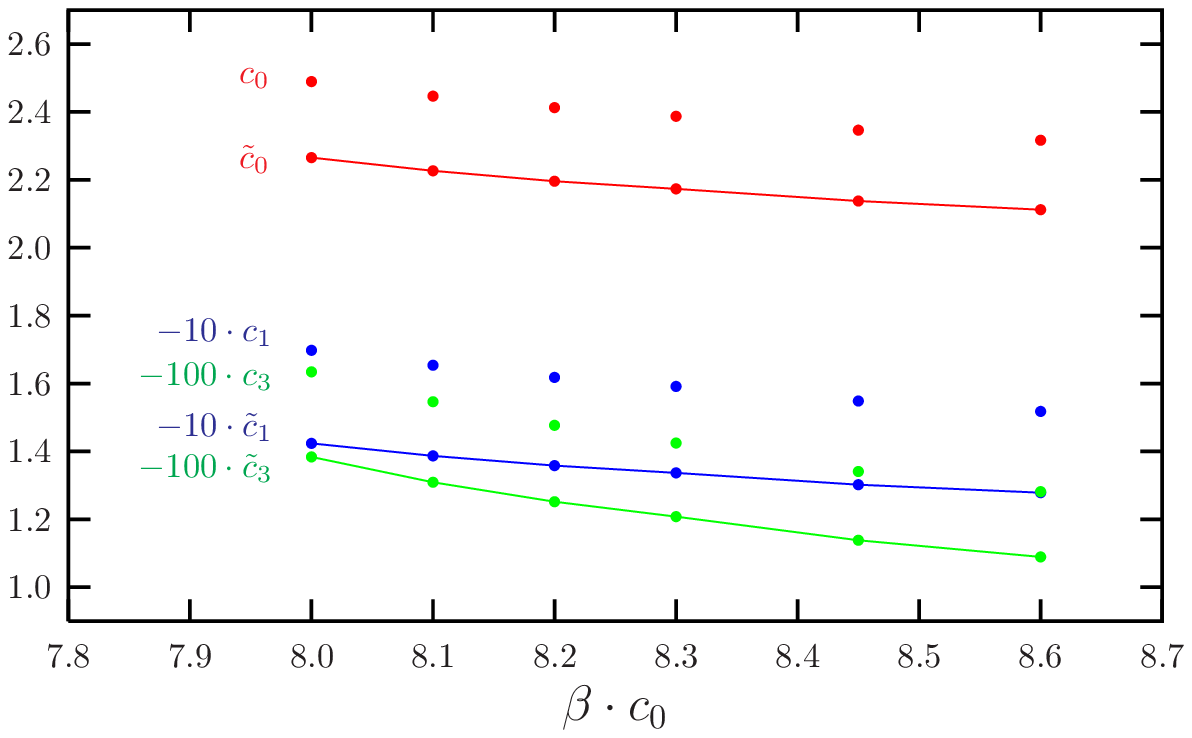,width=7.5truecm}
\vskip 1mm
\small{\label{fig5}Fig.5: Coe{f}f{i}cients $c_i$ and their dressed counter\-parts 
\ \~{c}$_i$
for different values of $\beta\,c_0 = 6\,c_0/g_0^2$ (TILW actions, $N=3$)} 
\end{center}
\end{minipage}
\vspace{3mm}
\begin{center}
\begin{minipage}{11cm}
\begin{center}
\begin{tabular}{cr@{}lr@{}lr@{}lr@{}l}
\hline
\multicolumn{1}{c}{$\beta=6/g_0^2$}&
\multicolumn{2}{c}{$c_0$}&
\multicolumn{2}{c}{$c_1$}&
\multicolumn{2}{c}{$\tilde{c}_0$} &
\multicolumn{2}{c}{$\tilde{c}_1$} \\
\hline
1.1636  &$\phantom{A^{3^2}}$5&.29078  &$\phantom{000}$-0&.53635  &
$\phantom{000}$3&.39826           &$\phantom{000}$-0&.22528    \\
0.6508  &12&.2688  &-1&.4086   &8&.8070            &-0&.7313     \\
\hline
\end{tabular}

\vskip 3mm
\small{TABLE I. Improved coe{f}f{i}cients\ \~{c}$_0$ and\ \~{c}$_1$ in the
  DBW2 action ($c_0$ and $c_1$ are obtained
{\it starting} from $\beta\,c_0=6.0$ and $6.3$)}
\end{center}
\end{minipage}
\end{center}

\section{Applications}
We now turn to two different
applications of cactus improvement: The additive mass renormalization
for clover fermions and the 1-loop renormalization of the axial and
vector currents using the overlap action. 
Both cases employ Symanzik improved gluons; hence, our
results are presented for various sets of Symanzik coe{f}f{i}cients. 

\subsection{Critical mass of clover fermions}
It is well known that an ultra-local discretization of the fermion
action without doubling breaks
chirality. Consequently, we must demand a zero renormalized fermion
mass, in order to ensure chiral symmetry while approaching the continuum
limit. For this purpose, the bare mass is additively renormalized from
its zero tree-level value to a critical value $dm$.

We calculate the 1-loop result for the critical mass $dm_{\rm 1-loop}$
using clover fermions and Symanzik improved gluons. The result is then
dressed with cactus diagrams in order to get the improved value
$dm_{\rm 1-loop}^{\rm dr}$. Details on the definition of $dm$ as well
as a 2-loop calculation of $dm$ with the same actions can be found in
Ref.~\cite{FP,PP,SCP}. The result of the 1-loop diagrams contributing
to $dm_{\rm 1-loop}$ can be written as a polynomial in the clover
parameter, and is independent of the number of fermion flavors $N_f$.
Some numerical values for $dm_{\rm 1-loop}$ corresponding to the plaquette
and Iwasaki actions ($N=3$) appear in Ref.~\cite{CPS}. 

Using the critical mass, one can evaluate the critical
hopping parameter, $\kappa_{\rm cr}\equiv{1/{(2\,dm + 8\,r)}}$
($r$ is the Wilson parameter). Estimates of $\kappa_{\rm cr}$ from
numerical simulations exist in the
literature for the plaquette action \cite{Bowler,LSSWW} ($N_f=0$),
\cite{JS} ($N_f=2$), and also the
Iwasaki action \cite{Khan} ($N_f=2$). Perturbative (unimproved and dressed) and
non-perturbative results are listed in Table II for specific
values of $c_{\rm SW}$. It is clear that cactus dressing
leads to results for $\kappa_{\rm cr}$  which are much closer to values
obtained from simulations. 
\vspace{-2mm}
\medskip
\begin{center}
\begin{minipage}{15cm}
\begin{center}
\begin{tabular}{lclllr@{}llr@{}llr@{}llr@{}l}
\hline
\multicolumn{1}{l}{Action}&$N_f$&\phantom{aa}&
\multicolumn{1}{c}{$\beta$}&$\phantom{\biggl(\biggr)}$&
\multicolumn{2}{c}{$c_{\rm SW}$}&\phantom{aa}&
\multicolumn{2}{c}{$\kappa_{\rm cr, 1-loop}^{\phantom{\rm dr}}$} &\phantom{aa}&
\multicolumn{2}{c}{$\kappa_{\rm cr, 1-loop}^{\rm dr}$} &\phantom{aa}&
\multicolumn{2}{c}{$\kappa_{\rm cr}^{\rm non-pert}$} \\
\hline
Plaquette\phantom{aa}&0&&6.00&$\phantom{A^{3^2}}$&
                 1&.479        &&0&.1301      &&0&.1362    &&0&.1392\\
Plaquette&0&&6.00&&1&.769      &&0&.1275      &&0&.1337    &&0&.1353\\
Plaquette&2&&5.29&&1&.9192     &&0&.1262      &&0&.1353    &&0&.1373\\
Iwasaki  &2&&1.95&&1&.53       &&0&.1292      &&0&.1388    &&0&.1421\\
\hline
\end{tabular}

\vskip 3mm
\small{TABLE II. 1-loop results and non-perturbative values for
  $\kappa_{\rm cr}$}
\end{center}
\end{minipage}
\end{center}

\subsection{One-loop renormalization of fermionic currents}
As a second application of cactus improvement, we investigate the
renormalization constant $Z_V$ ($Z_A$) of the flavor non-singlet
vector (axial) current in 1-loop 
perturbation theory. Overlap fermions and Symanzik improved gluons
are employed. Bare 1-loop results for $Z_{V,A}$ have been computed in
the literature \cite{AFPV,HPRSS,IP}; they depend on the overlap
parameter $\rho$ ($0<\rho<2$). 

The renormalization constants $Z_V$ and $Z_A$ are equal \cite{AFPV}
when using the overlap action and in the
$\overline{MS}$ scheme. Table III of Ref.~\cite{CPS} gives the values of $Z_{V,A}$ and
$Z_{V,A}^{\rm dr}$ for different sets of the Symanzik
coe{f}f{i}cients, choosing $\rho=1.0$, $\rho=1.4$. The
dependence of $Z_{V,A}$ and $Z_{V,A}^{\rm dr}$ on the
overlap parameter $\rho$ is shown in Fig. 6, where we 
plot our results for three actions: Plaquette, Iwasaki and
TILW. Note that improvement is more apparent for
the case of the plaquette action. This was to be expected,
since improved gluon actions were constructed in a way as to reduce lattice
artifacts, in the first place. A comparison between our improved $Z_{V,A}$ values and some
non-perturbative estimates \cite{GHR}, shows that improvement
moves in the
right direction.

\begin{center}
\psfig{file=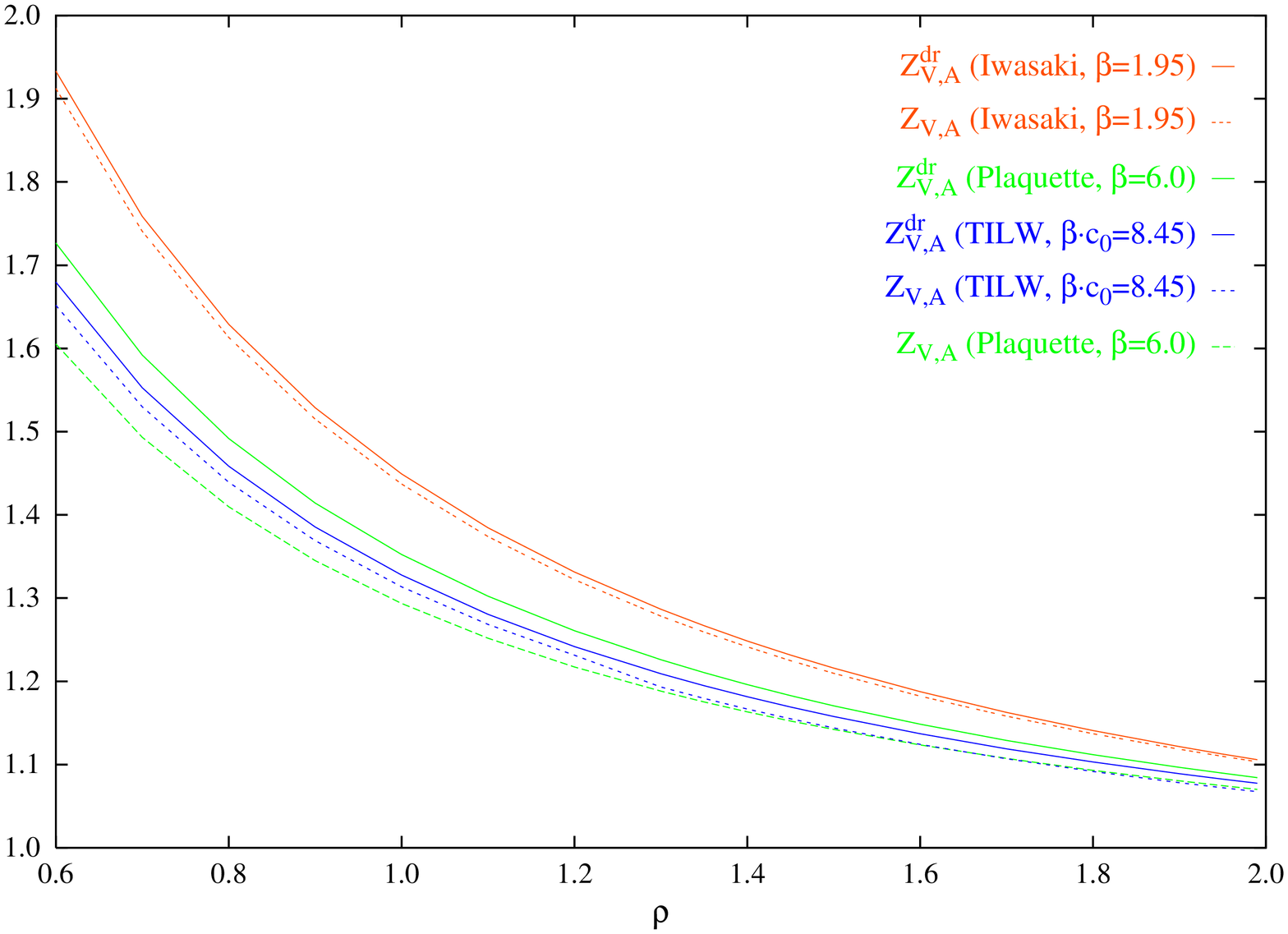,scale=0.45,angle=-90}
\vskip 2mm
\small{Fig. 6: Plots of $Z_{V,A}$ and $Z_{V,A}^{\rm dr}$ for the plaquette,
  Iwasaki and TILW actions. Labels have been placed in the same
  top-to-bottom order as their corresponding curves.}
\end{center}
{\bf Acknowledgements:} Work supported in part by the Research Promotion Foundation of Cyprus.


\end{document}